\documentstyle[fleqn,run2col,epsfig]{article}
\newcommand{\AmS}{{\protect\the\textfont2
  A\kern-.1667em\lower.5ex\hbox{M}\kern-.125emS}}

\title{Parton densities for heavy quarks}

\author{J. Smith\address{C.N. Yang Institute for \\
        Theoretical Physics, \\
        SUNY at Stony Brook , \\
        Stony Brook, NY 11794-3840}%
        \thanks{Work supported in part by the NSF grant PHY-9722101}}

\begin{document}

\begin{abstract}
We compare parton densities for heavy quarks.
\end{abstract}

\maketitle

Reactions with incoming heavy (c,b) quarks are often
calculated with heavy quark densities
just like those with incoming light mass (u,d,s) quarks
are calculated with light quark densities.
The heavy quark densities are derived 
within the framework of the so-called zero-mass 
variable flavor number scheme (ZM-VFNS). In this scheme these quarks 
are described by massless densities which are zero below 
a specific mass scale $\mu$. The latter depends on $m_c$ or $m_b$.
Let us call this scale the matching point.
Below it there are $n_f$ 
massless quarks described by $n_f$ massless densities.
Above it there are
$n_f + 1$ massless quarks described by $n_f + 1$ massless densities.
The latter densities are used to  
calculate processes with a hard scale $M \gg m_c, m_b$.
For example in the production of single top quarks via the weak process
$q_i + b \rightarrow  q_j + t$, where $q_i$, $q_j$ are light mass 
quarks in the proton/antiproton, one can argue that $M = m_t$
should be chosen as the large scale and $m_b$ can be neglected. 
Hence the incoming bottom quark
can be described by a massless bottom quark density.

\begin{figure}[h]
   \epsfig{file=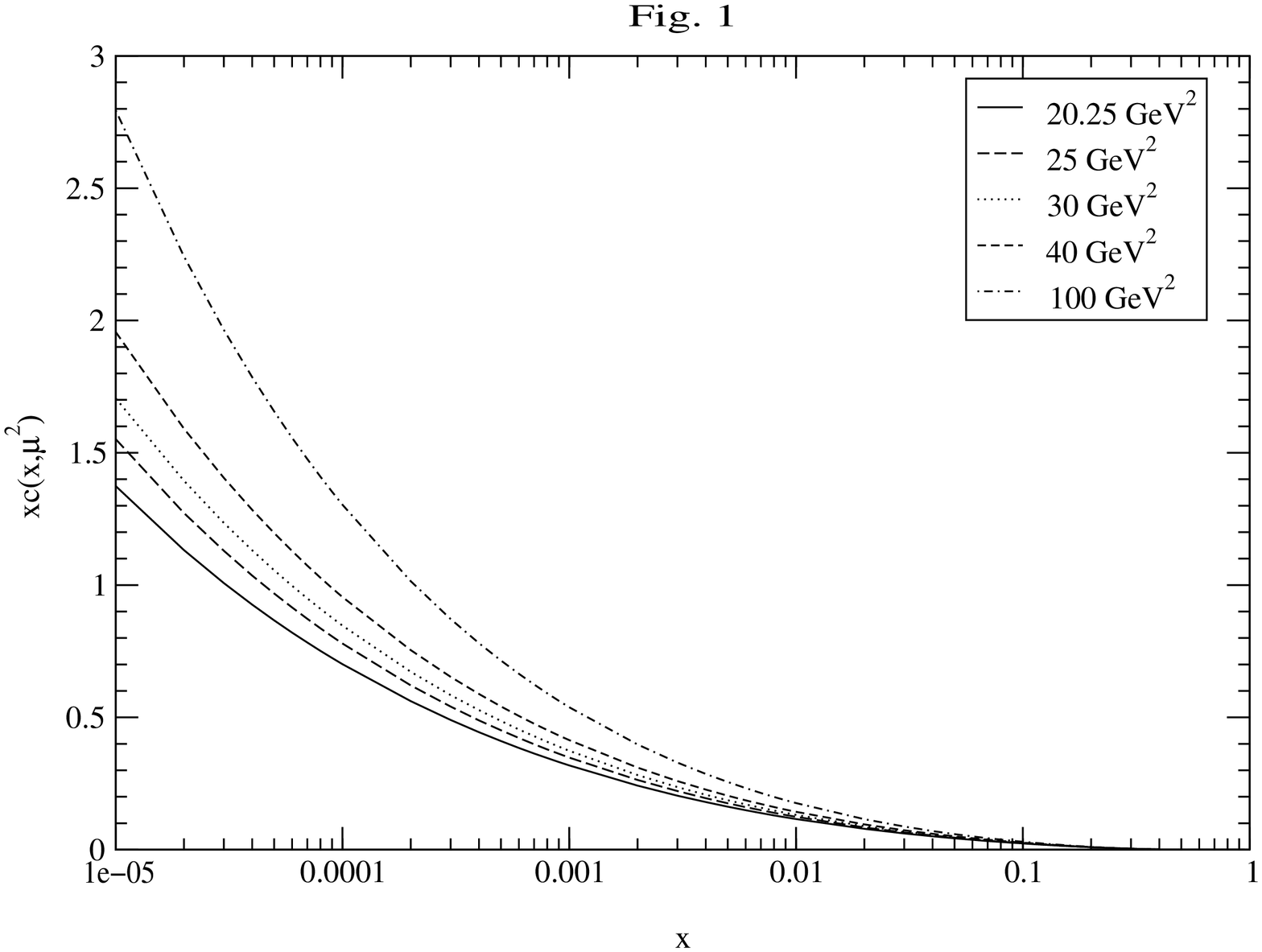,width=6cm,angle=270 }
\vspace{-0.8cm}
\caption{The charm quark density $xc_{\rm NNLO}(5,x,\mu^2)$ in the range
$10^{-5} < x < 1$ for $\mu^2 =$ 20.25, 25, 30, 40 and 100 in units
of $({\rm GeV/c}^2)^2$.}
\vspace{-0.4cm}
\end{figure}                                                              

The generation of these densities starts from the solution of
the evolution equations for $n_f$ massless quarks below the matching point.
At and above this point one solves the evolution equations for $n_f+1$
massless quarks. However in contrast to the parameterization
of the $x$-dependences of the light quarks and gluon at the initial
starting scale,
the $x$ dependence of the heavy quark 
density at the matching point is fixed. In perturbative QCD it is
defined by convolutions of the densities for the $n_f$ quarks 
and the gluon with specific operator matrix elements (OME's), 
which are now know up to $O(\alpha_s^2)$ \cite{bmsn1}. 
These matching conditions determine both the ZM-VFNS density
and the other light-mass quark and gluon densities at the matching points.
Then the evolution equations determine the new densities at larger
scales. The momentum sum rule is satisfied for the $n_f + 1$ quark
densities together with the corresponding gluon density.
\begin{figure}[h]
   \epsfig{file=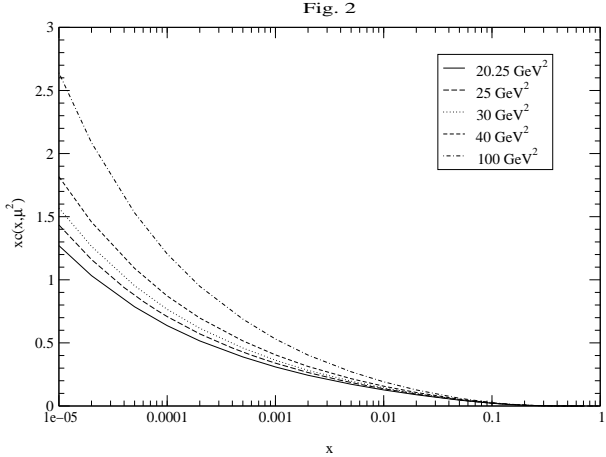,width=6cm,angle=270 }
\vspace{-0.5cm}
\caption{Same as Fig.1 for the NLO results from MRST98 set 1.}
\vspace{-0.4cm}
\end{figure}                                                              
\begin{figure}[h]
   \epsfig{file=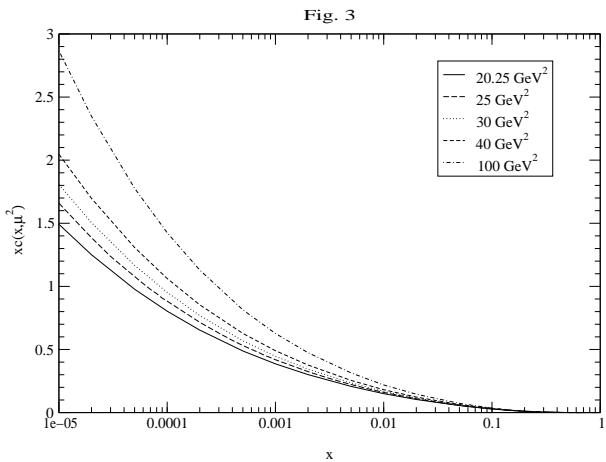,width=6cm,angle=270 }
\vspace{-0.5cm}
\caption{Same as Fig.1 for the NLO results from CTEQ5HQ.} 
\vspace{-0.4cm}
\end{figure}                                                              

Parton density sets contain densities for charm and bottom quarks, which
generally directly follow this approach or some modification of it. 
The latest CTEQ densities \cite{cteq5} 
use $O(\alpha_s)$ matching conditions. The $x$ 
dependencies of the heavy c and b-quark densities 
are zero at the matching points.
The MRST densities \cite{mrst98} have more complicated matching conditions 
designed so that the derivatives of the deep inelastic structure 
functions $F_2$ and $F_L$ with 
regard to $Q^2$ are continuous at the matching points.
Recently we have provided another set of ZM-VFNS densities \cite{cs1}, 
which are based on extending the GRV98 three-flavor densities in \cite{grv98}
to four and five-flavor sets. GRV give the formulae for their 
LO and NLO three flavor densities at very small scales. 
They never produced a c-quark density but advocated that charm quarks
should only exist in the final state of production reactions, which
should be calculated from NLO QCD with massive quarks as in \cite{lrsn}.
We have evolved their LO and NLO densities across the 
matching point $\mu = m_c$ with $O(\alpha_s^2)$
matching conditions to provide LO and NLO four-flavor densities containing 
massless c-quark densities. Then these LO and NLO densities were evolved 
between $\mu = m_c$ and $\mu = m_b$ with four-flavor LO and
NLO splitting functions. 
At this new matching point the LO and NLO four-flavor densities were then 
convoluted with the $O(\alpha_s^2)$ OME's to form five-flavor
sets containing massless b-quarks. These LO and NLO
densities were then evolved 
to higher scales with five-flavor LO and NLO splitting functions. 
Note that the $O(\alpha_s^2)$ matching conditions should
really be used with NNLO splitting functions to produce NNLO
density sets. However the latter splitting functions are not yet available, 
so we make the approximation of replacing the NNLO splitting functions
with NLO ones. 
\begin{figure}[h]
   \epsfig{file=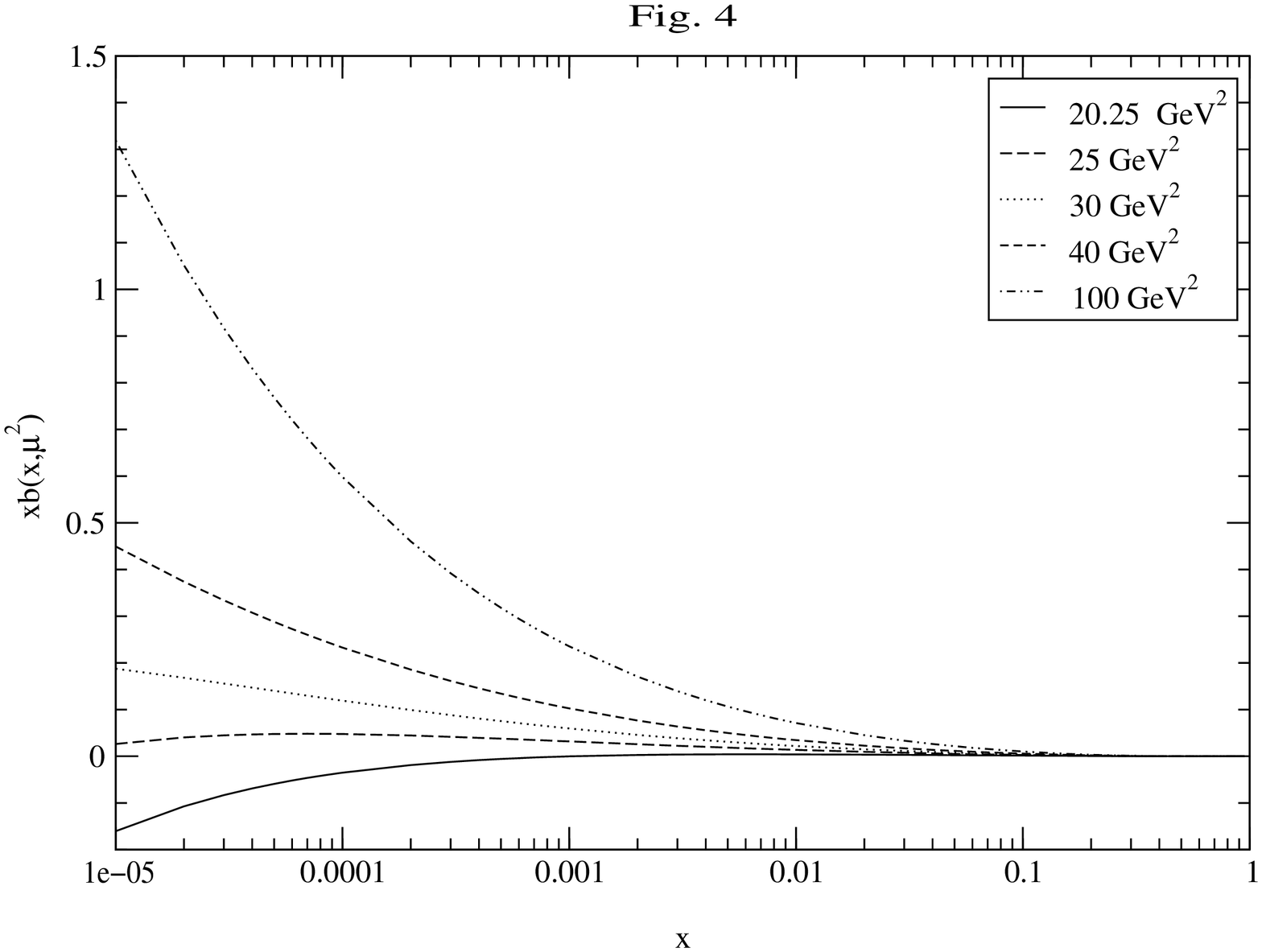,width=6cm,angle=270 }
\vspace{-0.8cm}
\caption{The bottom quark density $xb_{\rm NNLO}(5,x,\mu^2)$ in the range
$10^{-5} < x < 1$ for $\mu^2 =$ 20.25, 25, 30, 40 and 100 in units
of $({\rm GeV/c}^2)^2$.}
\vspace{-0.4cm}
\end{figure}                                                              

In this short report we would like to compare the charm and bottom 
quark densities in the CS, MRS and CTEQ sets.
We concentrate on the five-flavor densities, which are more important
for Tevatron physics. In the CS set they start at $\mu^2 = m_b^2 = 20.25$ 
${\rm GeV}^2$. At this scale the charm densities in the
CS, MRST98 (set 1) and CTEQ5HQ sets are shown in Figs.1,2,3
respectively. Since the CS charm density starts off negative for small $x$ at
$\mu^2 = m_c^2 = 1.96$ ${\rm GeV}^2$ it evolves less than the corresponding
CTEQ5HQ density. At larger $\mu^2$ all the CS curves in Fig.1 are below 
those for CTEQ5HQ in Fig.3 although the differences are small.
In general the CS c-quark densities
are more equal to those in the MRST (set 1) in Fig.2.

\begin{figure}[h]
   \epsfig{file=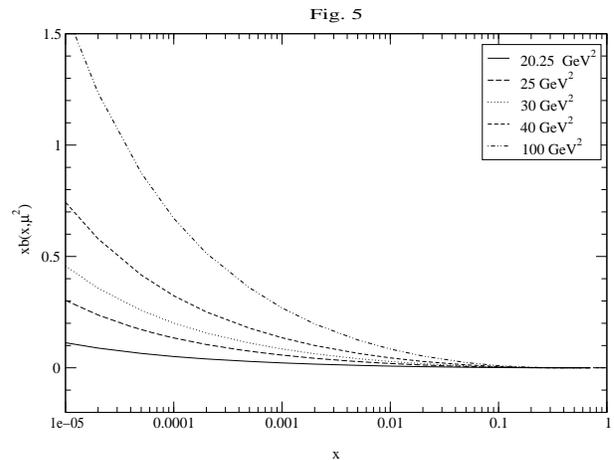,width=6cm,angle=270 }
\vspace{-0.5cm}
\caption{Same as Fig.4 for the NLO results from MRST98 set 1.}
\vspace{-0.4cm}
\end{figure}                                                              

At the matching point $\mu^2 = 20.25$ ${\rm GeV}^2$
the b-quark density also starts off negative at small $x$ 
as can be seen in Fig.4, which is a consequence of the explicit
form of the OME's in \cite{bmsn1}. At $O(\alpha_s^2)$ the OME's
have nonlogarithmic terms which do not vanish at the matching point
and yield a finite function in $x$, which is the boundary value
for the evolution of the b-quark density.
This negative start slows down the evolution of the b-quark density
at small $x$ as the scale $\mu^2$ increases. Hence the CS densities 
at small $x$ in Fig.4
are smaller than the MRST98 (set 1) densities in Fig.5 and the CTEQ5HQ 
densities in Fig.6 at the same values of $\mu^2$.
The differences between the sets are still small, of the order of five percent
at small $x$ and large $\mu^2$. Hence it should not really matter
which set is used to calculate cross sections for processes
involving incoming b-quarks at the Tevatron.

\begin{figure}[h]
   \epsfig{file=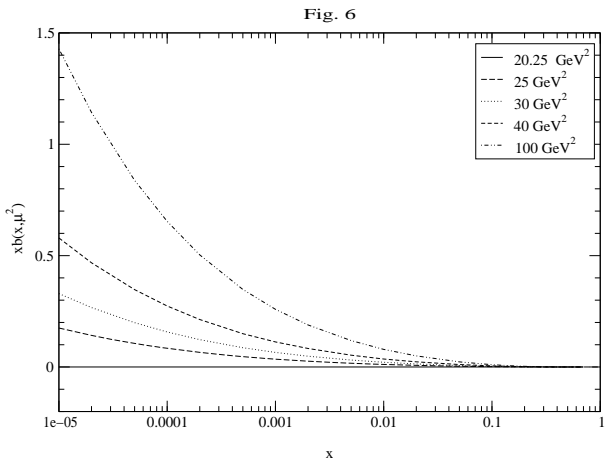,width=6cm,angle=270 }
\vspace{-0.5cm}
\caption{Same as Fig.4 for the NLO results from CTEQ5HQ.}
\vspace{-0.4cm}
\end{figure}                                                              

We suspect that the differences between these results 
for the heavy c and b-quark densities are primarily due to the
different gluon densities in the three sets rather to than the effects
of the different boundary conditions.
This could be checked theoretically if both LO and
NLO three-flavor sets were provided by MRST and CTEQ
at small scales. Then we could rerun our programs to generate
sets with $O(\alpha_s^2)$ boundary conditions. However these inputs are
not available. We note that CS uses the GRV98 LO and NLO gluon densities, 
which are rather steep in $x$ and generally
larger than the latter sets at the same values of $\mu^2$.
Since the discontinuous boundary conditions
suppress the charm and bottom densities at small $x$, they enhance the
the gluon densities in this same region (in order that the
momentum sum rules are satisfied).
Hence the GRV98 three flavour gluon densities and the 
CS four and five flavor gluon densities are generally significantly  
larger than those in MRST98 (set 1) and CTEQ5HQ. Unfortunately 
experimental data are not yet precise enough to decide which set is
the best one. We end by noting that all these densities are 
given in the $\overline{\rm MS}$ scheme.


\begin{thebibliography}{9}
%
\bibitem{bmsn1}
M. Buza, Y. Matiounine, J. Smith, W.L. van Neerven,
Eur. Phys. J. {\bf C1}, 301 (1998).
\bibitem{cteq5}
H.L. Lai, J. Huston, S. Kuhlmann, J. Morf\'{i}n, F. Olness, J. Owens,
J. Pumplin, W.K. Tung, hep-ph/9903282.
\bibitem{mrst98}
A.D. Martin, R.G. Roberts, W.J. Stirling and R. Thorne, 
Eur. Phys. J. {\bf C4}, 463 (1998).
\bibitem{cs1} 
A. Chuvakin, J. Smith, hep-ph/9911504, to be published in Phys. Rev. D.
\bibitem{grv98}   M. Gl\"uck, E. Reya and A. Vogt,
Eur. Phys. J.  {\bf C5}, 461 (1998).
\bibitem{lrsn}
E. Laenen, S. Riemersma, J. Smith and W.L. van Neerven,
Nucl. Phys. {\bf B392}, 162 (1993); ibid. 229 (1993);
S. Riemersma, J. Smith and W.L. van Neerven, Phys. Lett. {\bf B347},
43 (1995);
B.W. Harris and J. Smith, Nucl. Phys. {\bf B452}, 109 (1995).
%
\end{thebibliography}
\end{document}